\documentclass[twocolumn,showpacs,preprintnumbers,amsmath,amssymb,showkeys]{revtex4}

\usepackage{graphicx}
\usepackage{dcolumn}
\usepackage{bm}
\begin{document}

\title{Luttinger Sum Rule in Slave-Particle Theories }

\author{Ivana Mrkonji{\' c}}
\email{ivanam@phy.hr}
\author{Slaven Bari{\v s}i{\' c}}
 \affiliation{Department of Physics,
University of Zagreb, POB 331, 10002 Zagreb, Croatia}

\date{7 November 2002}

\begin{abstract}
The long-standing belief is that the mean-field-like decoupling
procedures applied to the slave-particle representations of the
problems with strong local interaction violate Luttinger sum rule.
The number of occupied resonant states is small and typically
equal to the deviation from the sum rule, shedding doubt on the
overall results. It is therefore illustrated, on the example of
the Emery model for the high-T$_c$ superconductors, that, through
the consistent application of the mean-field procedure to the
Hamiltonian and the propagators, the sum rule is restored and the
resonant band conserved. In addition to the resonant band, the
electron spectrum contains large number of occupied states at the
bare site-energy of the ion with strong repulsion. These results
can be straightforwardly generalized to the other similar
problems.
\end{abstract}
 \maketitle

The study of strongly correlated electron systems has been one of
the most active research fields of solid state physics for a long
time. In the present context, it is appropriate to mention
Friedel/Anderson and Kondo impurity problems, including the case
of Kondo lattice, the (extended) Hubbard models and their t-J
derivatives and the Emery model of the CuO$_2$ conducting planes
for the high-T$_c$ rare-earth oxides with its t-J limit. General
feature of these models is the appearance of the resonant state,
or the resonant band in the translationally invariant case, on the
top of a large fermionic background.

Many methods used in the treatment of strongly correlated systems
introduce auxiliary fermions and bosons
\cite{Barnes,Coleman,Read,Mil,Kot}. Although the transformation to
new particles is exact at the outset, it actually becomes useful
in approximations applied to the coupled fermion-boson system.
Especially simple situation occurs when only the on-site
interaction is retained and taken to infinity, because then only
one fermion and one boson field are sufficient. These two fields
are coupled and various decoupling schemes, such as mean-field
approximations with their Gaussian extensions or the local gauge
theories where the vertex corrections are ignored, lead to
representation of the Green function of the physical electron as
the product (or convolution) of the auxiliary fermion and boson
Green functions. This results in (satisfactory) appearance of the
resonant state or band. However, as the rule, the weight of the
resonance is small and of the order of the error in the Luttinger
sum rule associated with the calculation of the resonance
\cite{Rai,Feng,Tutis}. This sheds doubt on the ability of the
simple decoupling schemes to describe the resonant band, its
weight being of the order of the error.

The improvements were attempted by including the fermion-boson
interactions in high orders with vertex corrections ignored,
considering in particular the non-local string fields, with the
conclusion that the Luttinger sum rule cannot be restored
\cite{Feng}. Alternatively, the time dependence of the boson
fields was postulated unimportant in the calculation of the
physical particle propagation \cite{Cape}. This restores the
Luttinger sum rule, but results in the appearance of the
unphysical incoherent background in the wide energy range of the
spectral density.

A simple argument is hence given that calculating the physical
particle propagator consistently with the overall decoupling
procedure the Luttinger sum rule is recovered without eliminating
the resonant band. The large spectral background appears, but
localized in energy to the vicinity of the non-renormalized deep
level. To be specific, the reasoning is illustrated on the case of
the 2d Emery model in the slave-boson representation
\cite{Kotliar,Grilli,Castellani,Levin,Don,Capri}, treated in the
mean-field approximation, but the reasoning can be
straightforwardly extended to the other models in similar
decoupling schemes.

In the usual definition of the Emery model \cite{Emery} for the
CuO$_2$ plane where $\varepsilon_d$ and $\varepsilon_p$ are the
bare Cu and O site energies respectively, $t_0$ is the bare Cu-O
hopping, t' is the O-O hopping and $U_d$ is the Cu on-site Coulomb
repulsion. When $U_d \approx \infty$ double occupancy is forbidden
and the original fermion field $c_R$ (spin indices are not shown
for brevity) which annihilates the d-hole on the Cu site is
expressed in term of boson $b_R$ and fermion $f_R$ as
\begin{equation}
\label{e1} c_R=f_Rb{_R^\dagger}
 \end{equation}
 and $c{_R^\dagger}$ is given as its conjugate.

The Hilbert space associated with auxiliary fermions and bosons is
highly redundant, and the physical subspace is characterized by
the no-double occupancy condition
\begin{equation}
\label{e3} Q_R=f{_R^\dagger}f_R+b{_R^\dagger}b_R=1.
 \end{equation}

 Using Eq.(\ref{e1}) for copper sites,
 combined with the fermion fields $p_R$ associated with
 oxygen sites, the Emery Hamiltonian transforms to
\begin{eqnarray} \label{e2} H= \sum_{s,R}[\varepsilon_d n{_R^d}
+ \varepsilon_p \sum _{i=x,y}n{_{R,i}^p}+ \sum
_{i=x,y}t_0(f{_R^\dagger}b_Rp_{R,i}+ \nonumber \\ \sum_{ \delta
}f{_R^\dagger}b_Rp_{R+ \delta,i}+ h.c.)+
t'(p{_{R,x}^\dagger}p_{R,y}+ h.c.)],
 \end{eqnarray}
where R is a site index and $\delta$ denotes nearest neighbors.
Total number N of d and p fermions transforms analogously to H of
Eq.(\ref{e2}). The representations of N and H are however not
unique, as any operator equal to zero in $Q_R=1$ subspace, such as
$ \nu f{_R^\dagger}f_Rb{_R^\dagger}b_R$, $\nu$ arbitrary constant,
can be added to $n{_R^d}$. In other words, $n{_R^d}$ can be
represented as

\begin{equation}
\label{e4} n{_R^d}=n{_R^f}(b_Rb{_R^\dagger}+ \nu
b{_R^\dagger}b_R),
 \end{equation}
the expression in the brackets equal to one in $Q_R=1$ space.
Irrelevant in the exact calculations, the choice of $\nu $ in the
brackets becomes important for any approximate procedure not
confined to the $Q_R=1$ space. Then the only way to keep $n{_R^d}$
in N and H independent on the boson variables, as it is in the
exact calculations, is to choose
\begin{equation}
\label{e5} \nu = -1
 \end{equation}
in Eq.(\ref{e4}).

Noteworthy, $Q_R$ commutes with H and N and therefore they are
diagonalized in the  $Q_R=1$ subspace. In particular, the physical
ground state $|G \rangle$ is the state of the lowest energy in
that subspace. The mean-field slave-boson procedure (MFSB)
approximates $|G \rangle$ by the product of the ground state
$|G(b_0) \rangle$ of the displaced harmonic oscillator $b'=b-b_0$
and the ground state of the $n=1+ x$ free f-fermions
$|G_0(f)\rangle$
\begin{equation}
\label{e6} |G \rangle \approx |G_{MF} \rangle = |G(b_0)
 \rangle \times |G_0(f) \rangle.
 \end{equation}
$|G(b_0) \rangle$ has the property that $\langle b{_R^\dagger}b_R
\rangle= \langle b{_R^\dagger} \rangle ^2 = \langle b_R \rangle
^2= b{_0^2}$. $|G_{MF} \rangle$ has a component outside the
$Q_R=1$ subspace, which is minimized with the requirement that
$Q_R=1$ is satisfied at average
\begin{equation}
\label{e7} \langle G_{MF}|Q_R| G_{MF} \rangle =  \langle n^f
\rangle + b{^2_0}=1,
 \end{equation}
using Eq.(\ref{e3}). Even when $|G_{MF} \rangle$ is chosen to
satisfy Eq.(\ref{e7}), its component outside $Q_R=1$ subspace is
still present. This can be easily seen from the calculation of
$\langle n_d \rangle$ using Eq.(\ref{e4}), $ \langle n{_R^d}
\rangle= \langle n{_R^f} \rangle (1+ (\nu +1)b{^2_0}) $. From this
result, it is obvious that, in the presence of long-range order
$b_0 \neq 0$, the use of $|G_{MF} \rangle$ should be combined with
the choice $\nu =-1$ in Eq.(\ref{e4}) leading the equality
\begin{equation}
\label{e8}  \langle n{_R^d} \rangle =\langle n{_R^f} \rangle.
 \end{equation}

 Using further Eq.(\ref{e6}), boson field can be
averaged out from $H+ \lambda \sum Q_R$ of Eqs.(\ref{e2}) and
(\ref{e3}). Eq.(\ref{e8}), i.e. Eq.(\ref{e5}) is essential in
evaluating the local term. The result is the MF fermion
Hamiltonian which describes free f-fermions on the $CuO_2$ lattice
with the effective (renormalized) parameters of the
non-interacting Emery model $t=b_0t_0$ and
$\varepsilon_f=\varepsilon_d+ \lambda$, while t' remains
unchanged. On the other hand, when the fermion variables are
averaged out, the Hamiltonian $\lambda b'{_R^\dagger}b'_R$ of the
displaced $b'_R$ harmonic oscillator is obtained. Here again, the
choice $\nu = -1$ is important because it eliminates the $b'$
dependence from the local term of $H+ \lambda \sum Q_R$, i.e. it
leaves $\lambda$ to govern alone the energy of $b'$ boson.

Noting that the energy of the $b'$ boson field vanishes for
$|G(b_0) \rangle$, t (i.e. $b_0$) and $\lambda$ are found by
minimizing the ground state energy $E_0$ of the fermion MF
Hamiltonian at fixed number of fermions $n=1 + x$.

Time-dependent Gaussian fluctuations around MF saddle point are
considered in the next step. A quantity convenient to discuss for
such a purpose is the single particle propagator between the Cu
sites. According to Eq.(\ref{e1})
\begin{equation}
\label{e13} G{_{RR'}^d}(\tau)= (-i)\langle
G|Tf{_{R}^\dagger}(\tau)b_{R}(\tau) f_{R'}b{_{R'}^\dagger}|G
\rangle.
\end{equation}
As in the case of $n_d$ in Eq.(\ref{e4}), the choice of  operators
entering the autocorrelation function $G_{RR}$ in Eq.(\ref{e13})
is not unique, because in particular $\nu
f{_R^\dagger}f_Rb{_R^\dagger}b_R$ can
 be added to it without changing $G{^d_{RR}}$
\begin{equation}
\label{e14} G{_{RR}^d}(\tau)= (-i)\:b{_0^2}\: \langle
G|Tf{_R^\dagger}(\tau)f_{R}b_R(\tau)b{_R^\dagger}+ \nu
f{_R^\dagger}f_{R} b{_R^\dagger}b_R|G \rangle.
\end{equation}

The f-fermions and b'-bosons decouple, when the f b' coupling is
omitted in the MF approximation, i.e. $G^d$ decouples into the
product of fermion and boson propagators. Noting that to this
order the b' boson propagator is local

\begin{equation}
\label{e15} iG{_{RR'}^d}(\tau)= b{_0^2}\: \langle
G_0(f)|Tf{_R^\dagger}(\tau)f_{R'} |G_0(f) \rangle,\:\:\:\: R\neq
R',
\end{equation}
and
\begin{eqnarray}
\label{e16} iG{_{RR}^d}(\tau)= \langle G_0(f)|T
f{_R^\dagger}(\tau) f_R|G_0(f) \rangle \nonumber \\ \langle
G(b_0)|T b_R(\tau)b{_R^\dagger}|G(b_0) \rangle +
 \nu \langle n^f \rangle
b{^2_0},
\end{eqnarray}
where $\nu $ is kept explicit in order to follow its trace in the
following discussion.  Note that the operator in Eq.(\ref{e14})
reduces to $n^d$ of Eq.(\ref{e4}) at $\tau =0$. Consistently with
the MF choice $\nu = -1$, Eq.(\ref{e5}) of the MF procedure, this
choice should be repeated in Eq.(\ref{e16}) when $|G \rangle$ is
replaced by $|G_{MF} \rangle$ in order to satisfy sum rule
$\langle n^d \rangle =\langle n^f \rangle$ of Eq.(\ref{e8}).
General structure $\bar{\eta^2}-\bar{\eta}^2$ of Eq.(\ref{e16}) is
then obtained for the correlation function G$_{RR}$ when
long-time/long-range order ($b_0 \neq 0$) is present. This
structure justifies the time-independent choice of the $\nu
$-operator in Eq.(\ref{e14}).

Mean-field free-particle propagators appear in Eqs.(\ref{e15}) and
(\ref{e16}). The Fourier transform ($R=R'$ included) of the MF
$\nu =-1$ free-fermion propagator is

\begin{equation}
\label{e17} G^f(\omega, {\bf k})= \frac {|m_f({\bf k})|^2}{\omega
- \varepsilon ({\bf k})+ i \eta \: {\rm sign} (\varepsilon ({\bf
k})- \mu) },
\end{equation}
where $\varepsilon ({\bf k})$ is the dispersion of the free
f-fermion band and $|m_f({\bf k})|^2$ is the probability of
finding it on the Cu-site, both quantities obtained by MFSB. MF
$\nu =-1$ boson propagator $\langle G(b_0)|T b'(\tau) b'^\dagger
|G(b_0) \rangle$ in Eq.(\ref{e16}) is given by

\begin{equation}
\label{e18} D^{b'}({\omega})= \frac {1}{\omega - \lambda + i \eta
}
\end{equation}
because, as already mentioned, free dynamics of the boson b' is
local and determined at $\nu =-1$ by the frequency $\lambda$
(Cartesian gauge of the boson field is meant here). Combining
Eqs.(\ref{e16}), (\ref{e17}) and (\ref{e18}), $G{_{RR}^d}(\omega)$
reads
\begin{eqnarray}
\label{e19}G{_{RR}^d}(\omega)= \int \frac {d{\bf k}}{(2 \pi)^2}
\theta (\mu - \varepsilon ({\bf k})) \frac {|m_f({\bf
k})|^2}{\omega - (\varepsilon ({\bf k})- \lambda)- i \eta
}\nonumber \\+[G{_{RR}^f}(\omega)-i \pi \nu \langle n_f \rangle
\delta(\omega)]b{_0^2}
\end{eqnarray}
The first term in Eq.(\ref{e19}) describes \cite{Tutis} the local
excitation of the dispersionless level of the energy
$\varepsilon_d = \varepsilon_f - \lambda$. The number of states
associated with this level is given by the integration of $\pi
^{-1} {\rm Im}G^d(\omega)$ over frequency, and, as it can be
easily seen using Eq.(\ref{e19}), it is equal to $ \langle n^f
\rangle$.The excitation energy $\Delta_{pd}= \varepsilon_p -
\varepsilon_d$ of the dispersionless level at $\varepsilon_d$,
large when $\lambda$ is large, is observed in the high-energy
spectroscopies.

The second term in Eq.(\ref{e19}) together with Eq.(\ref{e15})
forms nearly half-filled resonant band described by
$b{_0^2}G^f(\omega, {\bf k})$. Band-dispersion of the resonant
band of the weight $b{_0^2}$ is obviously the one of the MF
solution \cite{Tutis}. The latter shows that the off-site dynamics
is determined by the small energy scales: $t<t_0$ (and $t'<t_0$ in
the spirit of the Emery model).These scales should be observed in
low frequency spectroscopies, such as ARPES.

The integrated contribution of the second term in Eq.(\ref{e19})
to $\pi ^{-1} {\rm Im}G^d(\omega)$ is $b{_0^2} \langle n^f
\rangle$. It corresponds to the contribution of the resonant band
$b{_0^2}G^f({\bf k}\omega)$ to $ \langle n^d \rangle$. Together
with the contribution $ \langle n^f \rangle$ of the first term and
$\nu b{_0^2} \langle n^f \rangle$ of the third, long-range order
term it gives $ \langle n^d \rangle = \langle n^f \rangle
(1+(1+\nu)b{_0^2})$, which for $\nu =-1$ reproduces the Luttinger
sum rule $ \langle n^d \rangle = \langle n^f \rangle$ of
Eq.(\ref{e8}). In other words, the long-range contribution cancels
exactly the contribution to $ \langle n^d \rangle$ of the resonant
band, without removing the resonant band itself.

Unfortunately, numerous previous calculations for the Emery,
Hubbard or Kondo problems \cite{Rai,Feng,Tutis,Cape,Kroha} amount
or are analogous to taking $\nu =-1$ in the MF procedure when
calculating the energy, but leaving \cite{Feng,Tutis,Cape} $\nu
=0$ in Eqs.(\ref{e14}-\ref{e19}) for the propagators. $\nu=0$
leads to $ \langle n^d \rangle = \langle n^f \rangle (1+b{_0^2})$,
i.e. to the breakdown of the Luttinger sum rule Eq.(\ref{e8}), the
discrepancy arising from the contribution of the resonant band. It
is therefore important to point out that by simple inclusion of
the long-range/long-time term in G$_{RR}$, by choosing $\nu =-1$
in Eqs.(\ref{e14}) and (\ref{e16}), solves the problem of the sum
rule without removing the resonant band.

Finally, it should be emphasized once again that the MF content of
the above argument is not essential for the restoration of the
Luttinger sum rule. The latter relies on two assumptions. The
first is that the ground state can be decoupled \cite{Feng} into
the product of boson and fermion states, as in Eq.(\ref{e6}). The
second assumption is that the long-time order is present in the
autocorrelation function i$G_{RR}$, leading to the inclusion of
the negative $\nu = -1$ time-independent term into Eq.(\ref{e16}).
The Hamiltonian has no explicit role in this argument, except that
the correlation functions appearing in the calculation of the
(ground state) energy have to be evaluated consistently with G.
The present argument applies therefore to the entire class of
Hamiltonians with large, local interactions \cite{Kroha}, amenable
to the slave-particle analysis.

\begin{acknowledgments}
  This work was supported by Croatian Ministry of Science under the
project 119-204.
\end{acknowledgments}

\end{document}